\documentstyle[aaspp4,psfig]{article}

\begin{document}

\def\gsim{\lower0.5ex\hbox{$\; \buildrel > \over \sim \;$}}
\def\lsim{\lower0.5ex\hbox{$\; \buildrel < \over \sim \;$}}
\def\rs{\mbox{$R_{\rm s}$}}
\def\rms{\mbox{$R_{\rm ms}$}}
\def\rh{\mbox{$R_{\rm h}$}}
\def\ro{\mbox{$R_0$}}
\def\ak{\mbox{$a_{\rm k}$}}
\def\nuu{\mbox{$\nu_{\rm u}$}}
\def\nuk{\mbox{$\nu_{\rm k}$}}
\def\num{\mbox{$\nu_{\rm m}$}}
\def\nub{\mbox{$\nu_{\rm b}$}}
\def\dnu{\mbox{$\Delta\nu$}}
\def\ms{\mbox{$M_{\odot}$}}

\def\be{\begin{equation}}
\def\ee{\end{equation}}

\title{Conversion of neutron stars to strange stars 
            as the central engine of $\gamma$-ray bursts} 

\author{ Ignazio Bombaci\altaffilmark{1}, and 
         Bhaskar Datta\altaffilmark{2,3} }

\altaffiltext{1}{ Dipartimento di Fisica, Universit\'a di Pisa, 
and INFN Sezione Pisa, via Buonarroti 2, I-56127, Italy;  
email: bombaci@pi.infn.it}
\altaffiltext{2}{Indian Institute of Astrophysics, Bangalore 560034, India}
\altaffiltext{3}{Raman Research Institute, Bangalore 560080, India  } 
\begin{abstract}
We study the conversion of a neutron star to a strange star as a 
possible energy source for $\gamma$-ray bursts.  
We use different recent models for the equation of state 
of neutron star matter and strange quark matter.  
We show that the total amount of energy liberated in the conversion is in 
the range (1---4)$\times 10^{53}$erg,  one order of magnitude 
larger than previous estimates, and in agreement with the energy 
required to power $\gamma$-ray burst sources at cosmological distances. 
\end{abstract}   

\section{Introduction}
 There is now compelling evidence to suggest that a substantial 
fraction of all gamma-ray bursts (GRBs) occur at cosmological distances 
(red shift $z\sim 1-3$). 
 In particular, the measured red shift $z=3.42$ for GRB~971214 
(Kulkarni et al. 1998), and $z\sim 1.6$ for GRB~990123 (Kulkarni et al. 1999) 
implies an energy release of $3 \times 10^{53}$erg and  
$3.4 \times 10^{54}$erg respectively,  in the $\gamma$-rays alone, assuming 
isotropic emission. 
The latter energy estimate could be substantially reduced if the energy 
emission is not isotropic, but displays a jet-like geometry (Dar 1998;  
Kulkarni et al. 1999).  
Models in which the burst is produced by a narrow jet are able to 
explain the complex temporal structure observed in many GRBs 
(Sari et al. 1997; 1999). 
In any case, a cosmological origin of GRBs leads to the conclusion of a huge 
energy output. Depending on the degree of burst beaming and on the efficiency 
of $\gamma$-ray production, the central engine powering these extraordinary 
events should be capable of releasing a total energy of a few $10^{53}$erg. 

 Many cosmological models for GRBs have been proposed. Among the most 
popular is the merging of two neutron stars (or a neutron star 
and a black hole) in a binary system (Paczynski 1998).   
Recent results (Janka \& Ruffert 1996) within this model, indicate that, 
even under the most favorable conditions, the energy provided by 
$\nu{\bar\nu}$ annihilation during the merger is too small by at least 
an order of magnitude, and more probably two or three orders of magnitude,  
to power typical GRBs at cosmological distances.  
An alternative model is the so--called "failed supernova" (Woosley 1993), 
or "hypernova" model (Paczynski 1998).  

 In the present work, we consider the conversion of a neutron star to a 
strange star (hereafter NS$\rightarrow$SS conversion) 
as a possible central engine for GRBs.  
In particular, we focus on the energetics of the NS$\rightarrow$SS conversion, 
and not on the mechanism by which $\gamma$-rays are produced.  
Previous estimate of the total energy $E^{conv}$ released in the 
NS$\rightarrow$SS conversion (Olinto 1987; Cheng \& Dai 1996) 
or in the conversion of a neutron star to hybrid star (Ma \& Xie 1996) 
gave $E^{conv}\sim 10^{52}$erg, too low to power GRBs at 
cosmological distances. These calculations did not include the various
details of the neutron star and strange star structural properties,
which go into the binding energy release considerations.
Here, we present accurate and systematic calculations of the total energy 
released in the NS$\rightarrow$SS conversion using different models for the 
equation of state (EOS)
of neutron star matter (NSM) and strange quark matter (SQM). 
We show that the total amount of energy liberated in the conversion is in 
the range  $E^{conv} = $(1---4)$\times 10^{53}$erg, in agreement with the 
energy required to power $\gamma$-ray burst sources at cosmological distances.

 The existence of strange stars (made up of degenerate u, d, and s quarks
in equilibrium with respect to the weak interactions) is allowable within 
uncertainties inherent in our present theoretical understanding of the 
physics of strongly interacting matter (Bodmer 1971; Witten 1984; 
Farhi \& Jaffe 1984).  
Thus, strange stars may exist in the universe, but until now, 
these have remained purely speculative entities. 
This situation changed in the last few years, thanks to the large amount of 
new observational data collected by the new generation of X-ray satellites.  
In fact, recent studies have shown that the compact objects associated with 
the X-ray bursters GRO J1744-28 (Cheng et al. 1998), SAX J1808.4-3658 
(Li et al. 1999a), and with the  X-ray pulsar Her X-1  (Dey et al. 1998) 
are good strange star candidates.  
Recently, Li et al. (1999b) have shown that the observed   
high-- and low--frequency  Quasi Periodic Oscillations (QPOs) 
in the atoll source 4U 1728-34 (M\'endez \& van der Klis 1999), 
are more consistent with a strange star compared to a neutron star,  
if the model of Osherovich \& Titarchuk (1999) correctly interprets 
the QPO phenomena.  

Originally, the idea that GRBs could be powered by the conversion of a neutron 
star to a strange star was proposed by Alcock {\it et al.} (1986) (see also 
Olinto (1987)), and recently reconsidered by other authors (Cheng \& Dai 1996). 
A similar model has been discussed by Ma and Xie (1996) for the conversion 
of a neutron star to a so--called hybrid star (a neutron star with a strange 
quark matter core).  

A number of different mechanisms have been proposed for the NS$\rightarrow$SS 
conversion.  All of them are based on the formation of a ``seed'' of SQM 
inside the neutron star. 
For example, 
(i) a seed of SQM enters in a NS and converts it to a SS (Olinto 1987). 
These seeds of SQM, according to Witten (1984),  are relics of 
the primordial quark--hadron phase transition microseconds after the Big Bang.   
(ii) a seed of SQM forms in the core of a neutron star as a result of a
phase transition from neutron star matter to deconfined strange quark matter 
(NSM$\rightarrow$SQM phase transition).  This could possibly happen when a 
neutron star (NS) is a member of a binary stellar system.   
The NS accretes matter from the companion star. The central density  of the NS 
increases and it may overcome the critical density for the NSM$\rightarrow$SQM 
phase transition. The NS is then converted to a SS.  
In the case of accretion induced conversion in a binary stellar system, 
the conversion rate has been estimated (Cheng \& Dai 1996) to be in the 
range (3--30)$\times 10^{-10}$ conversions per day per galaxy. 
This rate is consistent with the observed GRBs rate.   

However, once there is a seed of SQM inside a neutron star, it is possible 
to calculate the rate of growth (Olinto 1987; Horvath \& Benvenuto 1988). 
The SQM front absorbs neutrons, protons, and hyperons (if present), liberating 
their constituent quarks. Weak equilibrium is then re-established by the weak 
interactions. 
As shown by Horvath and Benvenuto (1988), the conversion of the whole star 
will then
occur in a very short time (detonation mode), in the range  1~ms -- 1~s, 
which is in agreement with the typical observed duration of GRBs. 
A detailed simulation of the conversion process is still lacking, and only  
rough estimates of the total energy liberated in the conversion have been made. 

As we show below, the dominant contribution to $E^{conv}$ 
arises from the internal energy  released in the conversion, {\it i.e.} 
in the NSM$\rightarrow$SQM phase transition.  
Moreover, the gravitational mass of the star will change during the 
conversion process, even under the assumption that the total number of baryons 
in the star is conserved.  
 
The total energy released in the  NS$\rightarrow$SS conversion is given by the 
difference between the total binding energy of the strange star BE(SS)  
and the total binding energy of the neutron star BE(NS)    
\be
                    E^{conv} = BE(SS) - BE(NS)
\ee
In the present work we assume that the baryonic mass $M_B$ of the compact 
object is conserved in the conversion process, {\it i.e}  
$M_B(SS) = M_B(NS) \equiv M_B$.  Then $E^{conv}$ is given in terms of    
the difference between the gravitational mass of the NS and SS:~   
$ E^{conv} = [M_G(NS) - M_G(SS)]c^2$.    

 In general the total binding energy for a compact object can be written 
   $  BE = BE_I + BE_G = (M_B - M_P)c^2  + (M_P - M_G)c^2$,          
where $BE_I$ and $BE_G$  denote the internal and gravitational binding 
energies respectively, and $M_P$ is the proper mass of the compact object 
defined as
\be
     M_P =   \int_0^R  dr
4\pi r^2 \Big[ 1 - {{2Gm(r)}\over{c^2 r}}\Big]^{-1/2}  \rho(r) ~, 
\label{eq:pmass}
\ee
$\rho(r)$ being the total mass--energy density and $m(r)$ the gravitational 
mass enclosed within a spherical volume of radius $r$. 
The proper mass is equal to the sum of the mass elements on the whole volume 
of the star; it includes the contributions of rest mass and internal energy 
(kinetic and interactions (other than gravitation)) of the constituents of 
the star. 
 
The total conversion energy can then be written as the sum of two contributions 
\be
       E^{conv} = E^{conv}_I  +  E^{conv}_G 
\ee
related to the internal and gravitational energy changes in the  conversion. 
These two contributions can be written as:
\be
     E^{conv}_I =  BE_I(SS) - BE_I(NS)  =  [M_P(NS) - M_P(SS)]c^2, 
\ee
\be
     E^{conv}_G =  BE_G(SS) - BE_G(NS)  =  
                   [M_P(SS) - M_G(SS) - M_P(NS) + M_G(NS)]c^2,    
\ee
and these can be evaluated solving the structural equations for   
non-rotating compact objects (Oppenheimer \& Volkoff 1939). 
To highlight the dependence of $E^{conv}$ upon the present uncertainties in 
the microphysics, we employed different models for the EOS of both NSM and 
SQM.    

Recently, a microscopic EOS of dense stellar matter has been calculated by 
Baldo et al. (1997), and used to compute the structure of static (Baldo et al. 
1997) as well as rapidly rotating neutron stars (Datta et al. 1998). 
In this model for the EOS, the neutron star core is composed of asymmetric 
nuclear matter in equilibrium, with respect to the weak interactions,   
with electrons and muons ($\beta$-stable matter).    
In particular, we consider their EOS based on the Argonne $v$14 nucleon-nucleon 
interaction implemented by nuclear three-body forces (hereafter BBB1 EOS).  

At the high densities expected in the core of a neutron star, additional 
baryonic states besides the neutron and the proton may be present, including 
the hyperons $\Lambda$, $\Sigma$, $\Xi$, $\Omega$, and the isospin 3/2 nucleon 
resonance $\Delta$.  The equation of state of this hyperonic matter is 
traditionally investigated in the framework of Lagrangian field theory in 
the mean field approximation (Glendenning 1985; Schaffner \& Mishustin 1996;  
Prakash et al. 1997). According to this model, the onset for hyperon formation 
in $\beta$-stable--charged neutral dense matter is about 2--3 times the normal 
nuclear matter density ($n_0 = 0.17$~fm$^{-3}$). The latter result has been 
confirmed by recent microscopic calculations based on the 
Brueckner-Hartree-Fock theory (Baldo et al. 1998).  
The appearance of hyperons, in general, gives a softening of the EOS with 
respect to the pure nucleonic case.   
In the present work we considered one of the EOS for hyperonic matter given 
in Prakash et al. (1997).  

For strange quark matter, we consider a simple EOS (Farhi and Jaffe 1984)  
based on the MIT bag model for hadrons. 
We begin with the case of massless non-interacting ($\alpha_c = 0$) quarks 
and with a bag constant $B = 60$ MeV/fm$^3$: we denote the corresponding 
EOS as $B60_0$.     
Next we consider a finite value for the mass of the strange quark within 
the same MIT bag model EOS. We take $m_s = 200$~MeV (and $m_u = m_d = 0$, 
$B = 60$ MeV/fm$^3$, and $\alpha_c = 0$; hereafter EOS $B60_{200}$).         
To investigate the effect of the bag constant on the energy released in 
the NS$\rightarrow$SS conversion, we take (almost) the largest 
possible value of $B$ for which SQM is still the ground state of strongly 
interacting matter, according to the so--called {\it strange matter hypothesis}  
(Witten 1984). For massless non-interacting quarks this gives 
$B = 90$ MeV/fm$^3$; we denote the corresponding EOS as $B90_{0}$.    

   Recently, Dey {\it et al.} (1998) derived an EOS for SQM using a 
different quark model with respect to the MIT bag model. 
The EOS by Dey {\it et al.} has asymptotic freedom built in, 
shows confinement at zero baryon density, deconfinement at high density, 
and, for an appropriate choice of the EOS parameters entering  
in the model, gives absolutely stable SQM according to the strange matter 
hypothesis. In this model, the quark interaction is described by  
a screened inter--quark vector potential originating from gluon exchange, 
and by a density-dependent scalar potential which restores the chiral 
symmetry at high density. 
The density-dependent scalar potential arises from the density dependence of 
the in-medium effective quark masses $M_q$, which, in the model by Dey et al. 
(1998), are taken to depend upon the baryon number density $n_B$ according to 
$M_q = m_q + 310 {\rm MeV} \times sech \big(\nu {{n_B}\over{n_0}}\big)$,   
where $n_0 $ is the normal nuclear matter density,  
$q (= u,d,s)$ is the flavor index, and $\nu$ is a parameter.   
The effective quark mass $M_q(n_B)$ goes from its constituent masses 
at zero density, to its current mass $m_q$,  as $n_B$ goes to infinity. 
Here we consider two different parameterizations of the EOS by Dey 
{\it et al.}, which correspond to a different choice for the parameter $\nu$. 
The  equation of state SS1 (SS2) corresponds to $\nu = 0.333$ ($\nu = 0.286$). 

\section{Results and conclusions}
To begin with, we fix as a ``standard'' EOS for neutron star matter 
the BBB1 EOS (Baldo et al. 1997), to explore how the energy budget in the  
NS$\rightarrow$SS conversion depends on the details of the EOS for strange 
quark matter. First we consider the $B60_0$ equation of state.    
The NS$\rightarrow$SS conversion based on this couple of EOSs, will be 
referred to as the  BBB1$\rightarrow B60_0$ conversion model. 
Similar notation will be employed according to the EOS of NSM and SQM.    
The total conversion energy, together with the partial contributions, is 
shown in the upper panel of Fig.~1.  As we can see, for $M_B$ larger than 
$\sim 1~M_\odot$ ({\it i.e.} values of the baryonic mass compatible with the 
measured neutron star gravitational masses) the total energy released in the 
NS$\rightarrow$SS conversion is in the range (1--3)$\times 10^{53}$erg, one 
order of magnitude larger than previous estimates (Olinto 1987; 
Cheng \& Dai 1996; Ma \& Xie 1996). Moreover, contrary to a common 
expectation, the gravitational conversion energy $E^{conv}_G$ is negative 
for this couple of EOSs.  
To make a more quantitative analysis, we consider a neutron star with 
a baryonic mass $M_B = 1.574 M_\odot$ (see tab. 1), which has a gravitational 
mass $M_G = 1.409 M_\odot$, a radius R(NS) = 11.0~km, and a gravitational 
binding energy  BE$_G$(NS) = 4.497 $\times 10^{53}$~erg. 
After conversion, the corresponding strange star has $M_G = 1.254 M_\odot$,  
R(SS) = 10.5~km, and BE$_G$(SS) = 3.061 $\times 10^{53}$~erg.    
The NS$\rightarrow$SS conversion is energetically possible in this case, 
thanks to the large amount of (internal) energy liberated in the 
NSM$\rightarrow$SQM phase transition.  

 Similar qualitative results for the total conversion energy are obtained for 
other choices of the two EOSs, but as we show below the magnitude of the two 
partial contributions are strongly dependent on the underlying EOS for NSM and 
SQM.  The total conversion energy for the BBB1$\rightarrow B60_{200}$ model is 
plotted in the lower left panel of Fig.~1. Comparing with the previous case, 
we notice that the strange quark mass produces a large modification of 
the conversion energy, which is reduced by a factor between 2--3 with respect 
to the case $m_s=0$. 
The bag constant $B$ has also a sizeable influence on the conversion energy. 
Increasing the value of $B$ reduces $E^{conv}$ and strongly modifies 
$E^{conv}_G$. This can be seen comparing the results for the 
BBB1$\rightarrow B60_0$ conversion model with those in the lower right 
panel of Fig.~1 for the BBB1$\rightarrow B90_0$ model. 
These results are a consequence of the sizeable effects of the strange 
quark mass and of the bag constant mainly on the internal binding energy 
$BE_I(SS)$ for strange stars (see {\it e.g.} Bombaci 1999).  
In fact, all strange stars configurations within the $B60_0$ EOS are 
self--bound objects ({\it i.e.} $BE_I(SS) > 0$).      
Strange star configurations within the $B90_0$ ($B60_{200}$) EOS are 
self--bound objects up to $M_G \sim 0.8~M_\odot$ ($M_G \sim 1.6~M_\odot$),  
to compare with the corresponding maximum gravitational 
mass $M_{max} = 1.60~M_\odot$ ($M_{max} = 1.75~M_\odot$).      

The results depicted in the two upper panels of Fig.~2  have been obtained 
using the EOS of Dey {\it et al.} (1998) for SQM, for two different choices  
of the parameter  $\nu$ which controls the rate at which chiral symmetry 
is restored to the quark masses at high density.   
For $\nu = 0.286$ (SS2), chiral symmetry is broken up to larger densities 
with respect to the case $\nu = 0.333$ (SS1). 
The parameter $\nu$ has a strong influence on the internal binding 
energy of the strange star.  
In fact, we found that strange stars within the SS2 (SS1) EOS are 
self--bound objects up to $M_G \sim 0.7~M_\odot$ ($M_G \sim 1.4~M_\odot$),  
to compare with the maximum gravitational mass 
$M_{max} = 1.33~M_\odot$ ($M_{max} = 1.44~M_\odot$). This effect is the 
main source for the differences in the calculated conversion energies for 
the two conversion models BBB1$\rightarrow$SS1 and BBB1$\rightarrow$SS2.  

The next step in our study is to consider a different neutron star matter 
EOS, which allows for the presence of hyperons in the neutron star core. 
We consider one of the EOS (hereafter Hyp) for hyperonic matter given in 
Prakash et al. (1997). 
In the two lower panels of Fig.~2 we plot the total conversion energy, 
together with the partial contributions, for the Hyp$\rightarrow B60_{0}$ 
and for the  Hyp$\rightarrow$SS1  conversion models. 
These results are in qualitative agreement with those reported in the  
previous figures. 

In table 1, we report the conversion energy, together with the partial 
contributions for the conversion of a neutron star with a gravitational mass 
$M_G(NS) \sim 1.4 M_\odot$ and for various conversion models. 

In the present work, we considered the conversion of a neutron star to a 
strange star as a possible energy source for $\gamma$-ray bursts. 
Our main focus was to perform an accurate calculation of the 
total released energy  compatible with our current understanding of the 
microphysics of strong interacting matter. 
We showed that the total amount of energy liberated in the 
NS$\rightarrow$SS conversion is in the range (1---4)$\times 10^{53}$erg,  
one order of magnitude larger than previous estimates, and in agreement 
with the energy required to power $\gamma$-ray burst sources at 
cosmological distances. 
\acknowledgements
One of the authors (I.B.) thanks Prof. J.E. Horvath for valuable 
discussions during the Quark Matter Conference in Torino.  



\vfil\eject
\begin{table*}
\caption{Conversion to strange star of a neutron star with 
$M_G\sim 1.4~M_\odot$,  for different EOSs for NSM and SQM. 
$M_B$ is the baryonic mass (which is conserved in the conversion process), 
$M_G(NS)$ is the neutron star gravitational mass, 
and $M_G(SS)$ is the gravitational mass of the corresponding strange star. 
All masses are in unit of the solar mass $M_\odot = 1.989\times 10^{33}$~g. 
$E^{conv}_G$, $E^{conv}_I$, and $E^{conv}$ are respectively the gravitational, 
internal, and total conversion energy, in unit of $10^{53}$~erg. } 
\vspace{.4 cm}
\label{tab1}
\begin{center}
\begin{tabular}{ccccccc}
\hline
NSM$\rightarrow$SQM      &  $M_B$        &   $M_G(NS)$    &  $M_G(SS)$  
                         &  $E^{conv}_G$ &   $E^{conv}_I$ &  $E^{conv}$ \\ 
\hline
BBB1$\rightarrow B60_0$         &  1.574  & 1.409  & 1.254 
                                & -1.436  & 4.215  & 2.779  \\
BBB1$\rightarrow B60_{200}$     &  1.574  & 1.409  & 1.340
                                & -0.677  & 1.920  & 1.243  \\
BBB1$\rightarrow B90_{0}$       &  1.573  & 1.409  & 1.343 
                                & -0.057  & 1.241  & 1.184  \\
BBB1$\rightarrow$SS1            &  1.558  & 1.397  & 1.235
                                &  0.580  & 2.308  & 2.888  \\
BBB1$\rightarrow$SS2            &  1.566  & 1.403  & 1.268
                                &  1.604  & 0.800  & 2.404  \\
Hyp$\rightarrow B60_0$          &  1.530  & 1.401  & 1.223
                                & -0.617  & 3.802  & 3.185  \\
Hyp$\rightarrow$SS1             &  1.530  & 1.401  & 1.217
                                &  1.291  & 2.002  & 3.293  \\
\hline
\end{tabular}
\end{center}
\end{table*}

\vfil\eject
\centerline{{\bf FIGURE CAPTIONS}}
\vskip 1.5cm
{\bf Fig. 1.~~}
The total energy liberated in the conversion of a neutron star 
to a strange star and the partial contributions from internal energy 
$E^{conv}_I$ (curves labeled ``Int'') and from gravitational energy 
$E^{conv}_G$ (curves labeled ``Gra'') as a function of $M_B$.  
See text for details on the equations of state for neutron star matter and 
strange quark matter
\vskip 1.5cm
{\bf Fig. 2.~~}
Same as in figure 1, but for different conversion models
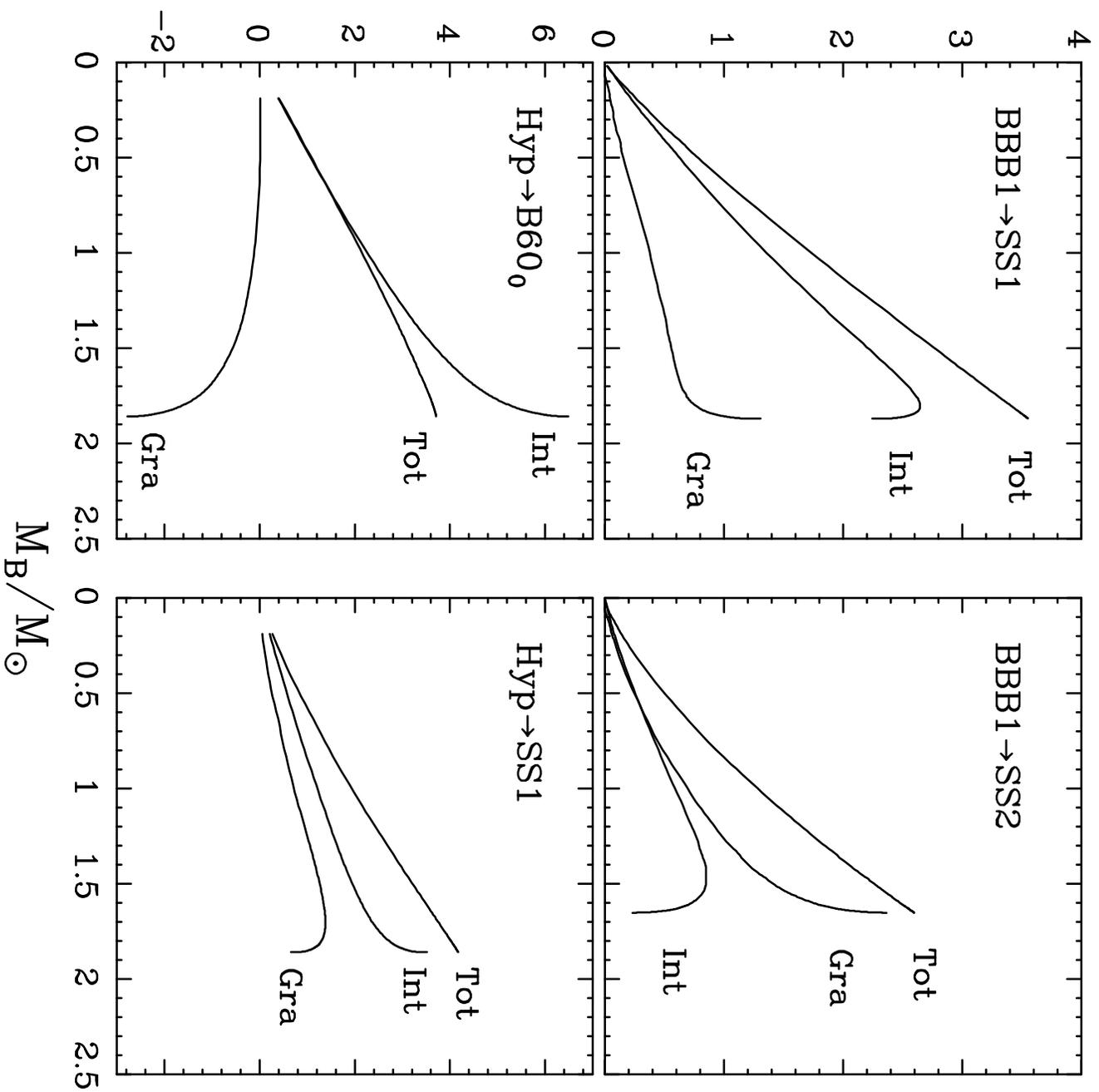
\end{document}